# Nanostructured antimony tin oxide synthesized via chemical precipitation method: its characterization and application in humidity sensing

B.C. Yadav[1,2]*, Rama Singh[1], Satyendra Singh[1], Ritesh Kumar[1] and Richa Srivastava[1]

[1]*Nanomaterials and Sensors Research Laboratory,*
*Department of Physics, University of Lucknow, Lucknow-226007, U.P., India*

[2]*Department of Applied Physics, School for Physical Sciences,*
*Babasaheb Bhimrao Ambedkar University, Lucknow-226025, U.P., India.*

* Email address: balchandra_yadav@rediffmail.com, Mobile: +919450094590

## Abstract

In present investigation we report the synthesis of antimony tin oxide nanoparticles via chemical precipitation method. The synthesized material was characterized using X-ray diffractometer, Scanning Electron Microscope, UV-visible absorption spectroscopy. XRD shows the crystalline nature of the synthesized material and the crystallite size was estimated by using Debye-Scherer equation and its minimum value was 3 nm. Pelletization of synthesized material was done using hydraulic press machine under uniform pressure of 616 MPa. Then the pellets were annealed at 200, 400 and 600°C. Further each pellet was put in humidity sensing chamber and corresponding variations in resistance with relative humidity (%RH) were measured. The average sensitivity was calculated by taking the average of all sensitivities ranging from 10 to 90% RH. The average sensitivity of the pellet annealed at 600°C was best among all the sensing pellets and was 2.18 KΩ/%RH. Results were reproducible ±84% after 2 months.

**Keywords:** Humidity sensor; sensing pellets; XRD; UV-visible analysis.

## 1. Introduction

Humidity sensors have found wide applications in industry production, production process control, environment monitoring, storage, and electrical applications. Therefore the research devoted to developments of new materials for sensor devices are gaining more and more attention. Recently, much attention has been focused on the modification of metal oxide by doping or substituting with special atoms [1]. Among these doping systems, antimony-doped tin dioxide (ATO) has attracted considerable attention owing to its potential applications as gas sensor, humidity sensor, solar battery, transparent electrode, electricity-conducting coatings and so on [2−3]. Therefore, different methods including solid blend, chemistry co-deposition, sol-gel, metal alkoxide hydroxylation and hydrothermal technologies have been proposed to prepare ATO [4]. Tin dioxide ($SnO_2$), an important n-type semiconductor with a wide band gap ($E_g$ = 3.6 eV), exhibits excellent optical, electrical and chemical properties and high thermal stability. In recent years, doped $SnO_2$ and $SnO_2$-based materials, such as Sb-doped $SnO_2$, Mn-doped $SnO_2$, $Zn_2SnO_4$, $Cd_2SnO_4$ and so on, have been extensively studied due to their special optical and electrical properties. During the past few years, these materials have been prepared by many techniques such as sol-gel [5], simple thermal evaporation [6], thermal CVD [7], hydrothermal methods [8-11] and other methods. On the other hand, the chemical and physical properties of these materials also depend on the sizes and shapes of particles. Synthesis of nanomaterials is very important step for humidity sensing applications as it decrease crystallite size and increase surface area of the sample. Finally, the nanocrystals are very attractive as primary building units for assembling nanostructured materials with defined porous architectures, which are promising as large-surface-area transparent electrodes for efficient optoelectronic devices. In porous ceramics, the surface and open pores tend to collect water vapour and gases through chemical and physical adsorption and through condensation [12-16]. Especially in semiconductor ceramics, electrical properties are largely related to the grain size and the pore size distribution of the open pores. In recent years, major domestic applications of humidity sensors have been the automatic humidity control in air conditioners and the automatic cooking in microwave ovens. In chemical and alimentary industries and in the production of electronic devices, rapid and accurate humidity control is of critical importance. The polymeric and ceramic sensors work utilizing the mechanism of the chemical and physical adsorption of water vapour on the surface of the material in a quantity that is proportional to the relative humidity of the surrounding environment. The adsorption generates a consequent variation of the resistance.

In the present investigation, we have synthesized $Sn_{.918}Sb_{.109}O_2$ by chemical precipitation method. The characterization of the synthesized material was done by using various techniques such as X-ray powder diffraction, Scanning electron microscopy and UV-visible absorption Spectroscopy. Further the pellets were made by using hydraulic press under uniform pressure of 616 MPa. These pellets were exposed with humidity in a specially designed humidity chamber. The average sensitivity of the pellet annealed at 600°C was best among all the sensing pellets.

## 2. Experimental procedure

All chemical reagents used for the preparation of $Sn_{.918}Sb_{.109}O_2$ were of AR grade. In the typical synthesis 5-10 ml of hydrochloric acid was added in double distilled water. At the moment $SnCl_2 \cdot 2H_2O$ and $SbCl_3$ in molar ratio 1:4 was mixed to above solution. Further ammonium hydroxide was added drop by drop to the above mixed solution. Small amount (5-10 ml) of ply-ethylene glycol was added as capping agent. Vigorous magnetic stirring was done for 18-24 h to ensure complete and intimate reaction between the various components. The product was dried 5-8 h at 80°C in an oven and calcined at 400°C for 3 h, resulting in complete crystallization to obtain $Sn_{.918}Sb_{.109}O_2$ into powder form. Then it was examined with wide angle XRD analysis (X-Pert, PRO PANalytical XRD system, Netherland). The crystallite size of powdered material was calculated from the X-ray line broadening, using the Debye-Scherer's equation. The pellets of the powder were made by hydraulic press machine (M. B. Instruments, Delhi, India) under a uniaxial pressure of 616 MPa at ambient temperature. Each pellet was annealed at 200°C, 400°C and 600°C and put within an Ag-pellet-Ag electrode system and exposed with humidity in a specially designed humidity chamber. Potassium sulphate solution in a dish was put within the chamber to increase the humidity from 10% to 95% in the chamber i.e. it acts as a humidifier whereas potassium hydroxide solution acts as a dehumidifier i.e. to decrease humidity from 95% to 10%. Measurement of electrical resistance was done using Keithley Electrometer (Model: 6514A). Then humidity sensing of each pellet for different annealing temperature was investigated.

## 3. Material Characterization

### 3.1 Scanning Electron Microscopy

Surface morphology of the sensing pellet annealed at 600°C was investigated using Scanning electron microscopy (SEM, LEO-0430, Cambridge). Figure 1 shows the SEM image of sensing pellet. Surface morphology of sensing material reveals that most particles are spherical in shape having large no of pores which is a promising consequence for good sensitivity. Higher porosity increases surface to volume ratio and hence helps in getting good sensitivity.

### 3.2 X-Ray Diffraction

Structural characterization and phase identification were performed by X-ray Diffractometer (X-Pert, PRO PANalytical XRD system, Netherland) with CuK$_α$ radiation as source having wavelength 1.54609 Å from 2θ = 20 to 100° with step size 0.1400° was done to investigate the phase formation in the composite. Figure 2 shows the XRD pattern of Sn$_{.918}$Sb$_{.109}$O$_2$. The XRD patterns indicate that tin antimony oxide nanoparticles have a tetragonal structure. Crystallite size of powder was calculated using Debye- Scherer formula which is as follows:

$$D = \frac{K\lambda}{\beta \cos\theta}$$

Where β is the full width at half maximum (FWHM) of the peak, λ is X-ray wavelength, θ is the Bragg angle and K = 0.94, a dimensionless constant. The peak centred at 2θ = 33.595° having highest intensity is assigned to tetragonal crystalline tin antimony oxide reflection (110) having d spacing 3.34978 Å and FWHM value as 3°. The crystallite size corresponding to this peak was found smallest (3 nm) among all peaks. Average crystallite size was found 6 nm.

### 3.3 UV–Visible absorption spectroscopy

Optical characterization was done by using UV-visible spectrophotometer (Varian, Carry-50Bio). UV–visible absorption spectroscopy is a useful technique for characterizing optical and electronic properties of different materials such as thin films, filters and pigments. It measures the percentage of radiation in the ultra-violet (200–400 nm) and visible (400–800 nm) regions that is absorbed at each wavelength. Measurement of the electronic band gap of semiconductor or films is carried out using the data obtained by spectrophotometer. There is a sharp increase in absorption at energies close to the band gap that manifests itself as an absorption edge in the UV–visible absorption spectra. The absorption spectra of tin antimony oxide nanoparticles obtained in the UV-visible region show absorption edge at 260 nm. The corresponding band gap was found 4.62 eV.

## 4. Humidity Sensing Mechanism

The basis of sensing is the change in electrical resistivity of semiconducting materials when exposed in humid environments. Sensing elements having large pore volume, large surface area, and suitable pore size distribution are the key elements for humidity sensing.

Water is absorbed on porous materials when they were exposed to humidity leading to an increase of their electrical conductivity [17-18]. Water adsorption in the internal surfaces of porous materials is important for electrical conduction, being dominant for low relative humidity. Adsorbed water condensed on the surface of materials and protons were conducted in form of aquatic layers. The larger is the surface area, larger is the content of adsorbed water, and consequently larger is the density of charge carriers usually protons. Physical adsorption and capillarity condensation of water molecules at the internal surfaces of the porous samples promote in an increase of the charge carrier concentration with a decrease of the electrical resistivity [19]. Water adsorption in the internal surfaces of the porous materials is important for the electrical conduction, being dominant for low relative humidity. In humid atmosphere, the reaction between the semiconducting material and water molecules is found to result in a change (increase or decrease) in resistivity. The mechanism involved is the physical or chemical adsorption of the gas molecules on the surface, leading either to a change in the carrier concentration, and hence the resistance by the exchange of electrons (acceptance or donation) between the adsorbed species and the material or to a change in the resistance due to a surface conduction mechanism involving the adsorbed molecules. The process of chemisorptions occurs at very low humidity levels, and is unaffected by further changes in humidity. However an increase in humidity makes the water molecules physisorb onto this hydroxyl layer. At higher humidity levels, the number of physisorbed layers increases allowing each water molecule to be singly bonded to a hydroxyl group, and proton hopping between adjacent water molecules in the continuous water layer takes place. The conduction process is the same as that of pure water and is called Grotthuss chain reaction [20].

## 5. Results and discussion

The semiconducting humidity sensors usually work on the principle of amount of water adsorbed on the pellet surface. Adsorption generates the variation in electrical resistance with humidity. Curve (a) of Figure 4 shows humidity sensing characteristic of sensing pellet at room temperature, curve (b) shows humidity sensing of pellet annealed at 200°C, similarly curves (c) and (d) show humidity sensing of the pellet annealed at 400 and 600 ºC respectively. Curve (a) of Figure 4 shows that resistance decreases sharply in the lower humidity range, i.e. from 10-20 %RH and further at higher humidity there is a continuous slow decrement in resistance. Curve (b) of Figure 4 shows that resistance decreases sharply up to 25 %RH and then linearly up to 65 %RH and further slow decrement was observed. Curve (c) of Figure 4 shows sharp decrease in resistance up to 40 %RH, following by linear decrease up to 65 %RH and then continuous decrease for higher humidity. Curve (d) of Figure 4 shows that resistance decreases sharply up to 35 %RH and there is linear decrease for high range of humidity up to 80 %RH, further continuous decreases. Sensitivity of a humidity sensor can be defined as the change in resistance (ΔR) of sensing element per unit change in relative humidity (%RH) i.e.

$$S = \frac{\Delta R}{\Delta RH\%} \, K\Omega/\%RH$$

Average sensitivity is calculated by taking the average of all sensitivities ranging from 10 to 90 %RH. Average sensitivity of the pellet sensor were found 0.1361 KΩ/%RH, 0.651 KΩ/%RH, 1.462 KΩ/%RH and 2.18 KΩ/%RH respectively for room temperature, annealed at 200, 400 and 600 °C. It has been observed that as the annealing temperature increases sensitivity of the sensing material increases and it becomes maximum for pellet annealed at 600°C. As more active sites are available for the pellet annealed at 600°C there was more change in resistance resulting in more sensitivity which is significant for sensor fabrication.

Figures 5, 6, 7 and 8 shows hysteresis curves for sensing pellets at room temperature, annealed at 200, 400 and 600°C respectively and these curves show 66%, 40%, 27% and 38% hysteresis respectively at room temperature, annealed at 200, 400 and 600°C respectively. After studying the humidity sensing properties of sensing elements, they have been kept in laboratory and their humidity characteristics were regularly monitored. To check the reproducibility, the sensing property of the sensing elements was again examined in the humidity control chamber after two months. Figure 9 shows the reproducibility of pellet annealed at 600°C. The sensor was reproducible up to ±84% after two months. Figure 10 shows variations in average sensitivity with different temperatures. It illustrates the pellet annealed at 600°C have maximum sensitivity in comparison to other temperature.

## 6. Conclusion

Nanostructured antimony tin oxide was successfully synthesized via chemical precipitation method. Minimum crystallite size was found 3 nm. Surface morphology of pellet annealed at 600°C shows that most particles are spherical in shape leaving most space as pores and hence it was most sensitive among all. Maximum average sensitivity for pellet annealed at 600°C was 2.18 KΩ/%RH which is best in comparison to others. Pellet annealed at 600°C have reproducibility ±84% after two months.

**Acknowledgement**

Authors are grateful to Department of Science and Technology, Government of India for SERC-FAST TRACK, Project SR/FTP/PS-21/2009.

**Figures Caption:**

**Figure 1** SEM image of $Sn_{.918}Sb_{.109}O_2$ pellet annealed at 600°C.

**Figure 2** XRD pattern of sensing material.

**Figure 3** UV-visible absorption of Antimony Tin Oxide.

**Figure 4** Variation in resistance with relative humidity.

**Figure 5** Hysteresis behaviour of $Sn_{.918}Sb_{.109}O_2$ pellet at room temperature.

**Figure 6** Hysteresis behaviour of $Sn_{.918}Sb_{.109}O_2$ pellet annealed at 200°C.

**Figure 7** Hysteresis behaviour of $Sn_{.918}Sb_{.109}O_2$ pellet annealed at 400°C.

**Figure 8** Hysteresis behaviour of $Sn_{.918}Sb_{.109}O_2$ pellet annealed at 600°C.

**Figure 9** Ageing effect of $Sn_{.918}Sb_{.109}O_2$ pellet annealed at 600°C.

**Figure 10** Variations in average sensitivity with different temperatures.

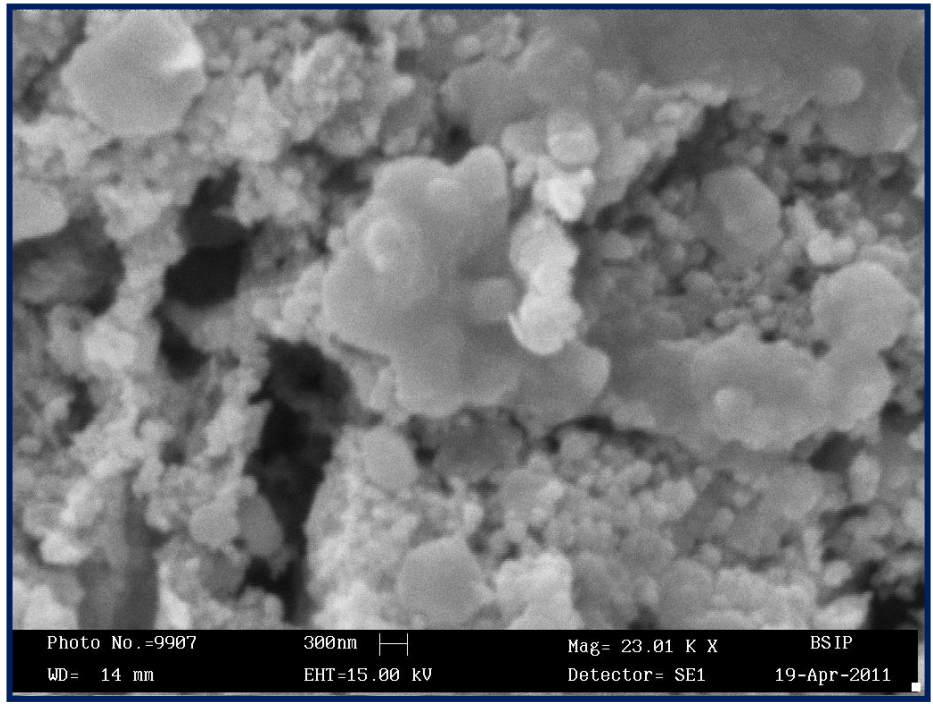

**Figure 1** SEM image of $Sn_{.918}Sb_{.109}O_2$ pellet annealed at 600°C.

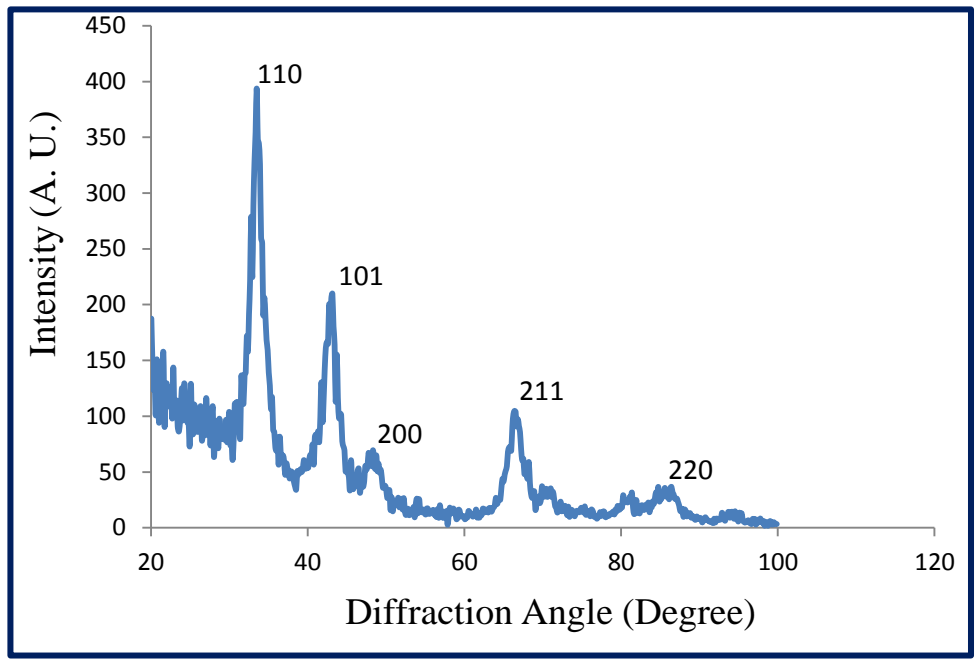

**Figure 2** XRD pattern of sensing material.

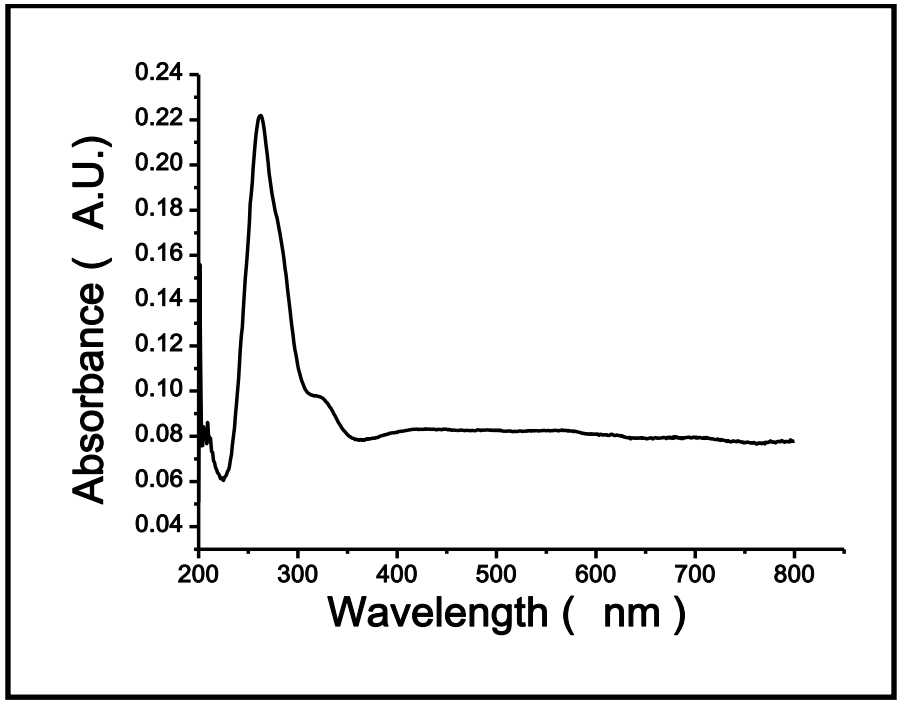

**Figure 3** UV-visible absorption of Antimony Tin Oxide.

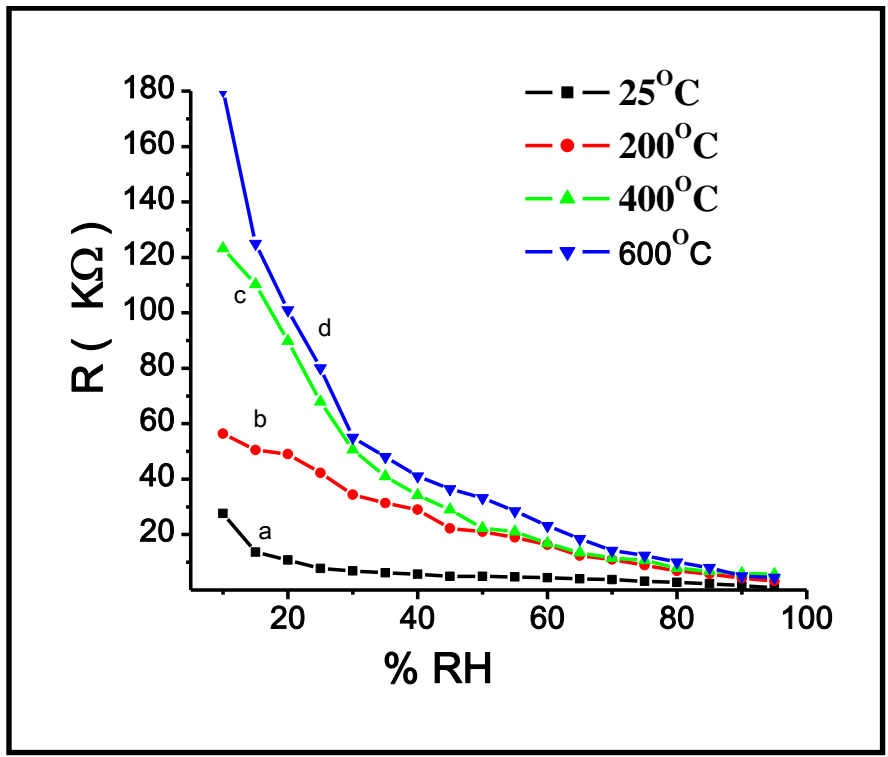

**Figure 4** Variation in resistance with relative humidity.

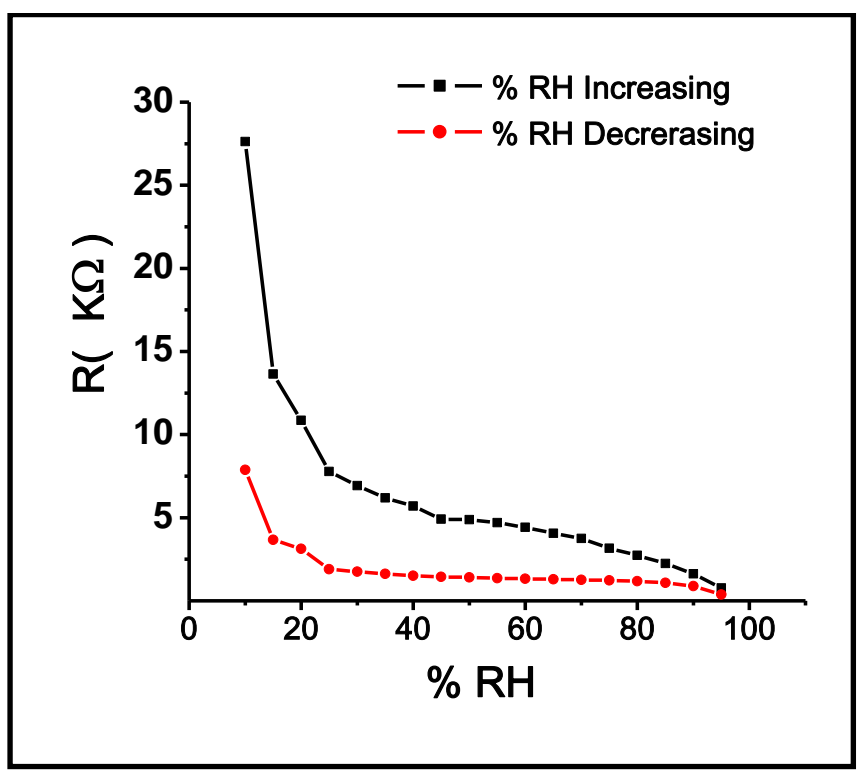

**Figure 5** Hysteresis behaviour of $Sn_{.918}Sb_{.109}O_2$ pellet at room temperature.

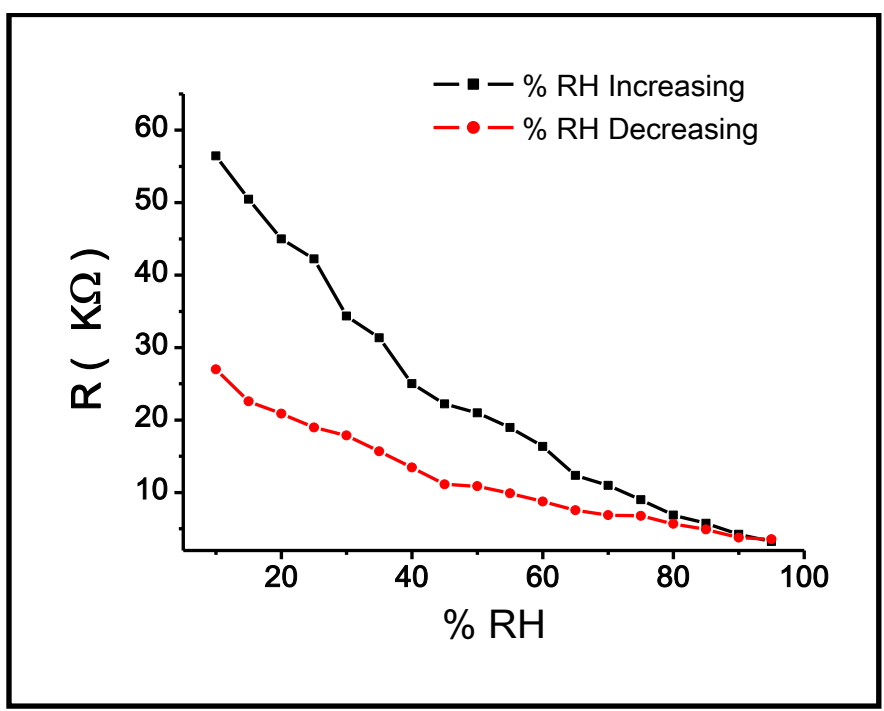

**Figure 6** Hysteresis behaviour of $Sn_{.918}Sb_{.109}O_2$ pellet annealed at 200ºC.

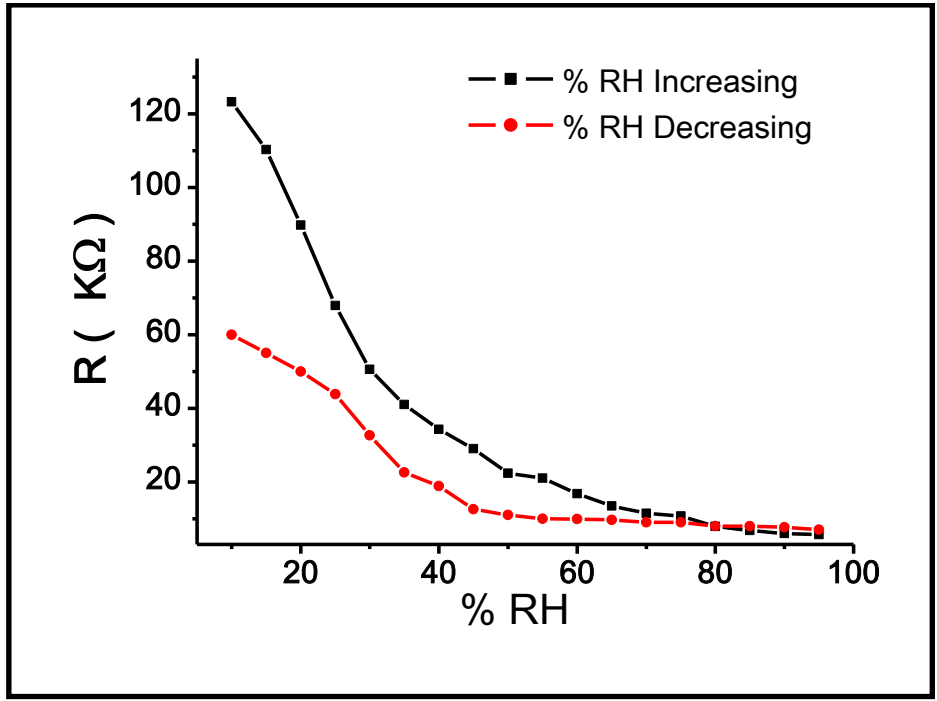

**Figure 7** Hysteresis behaviour of Sn$_{.918}$Sb$_{.109}$O$_2$ pellet annealed at 400ºC.

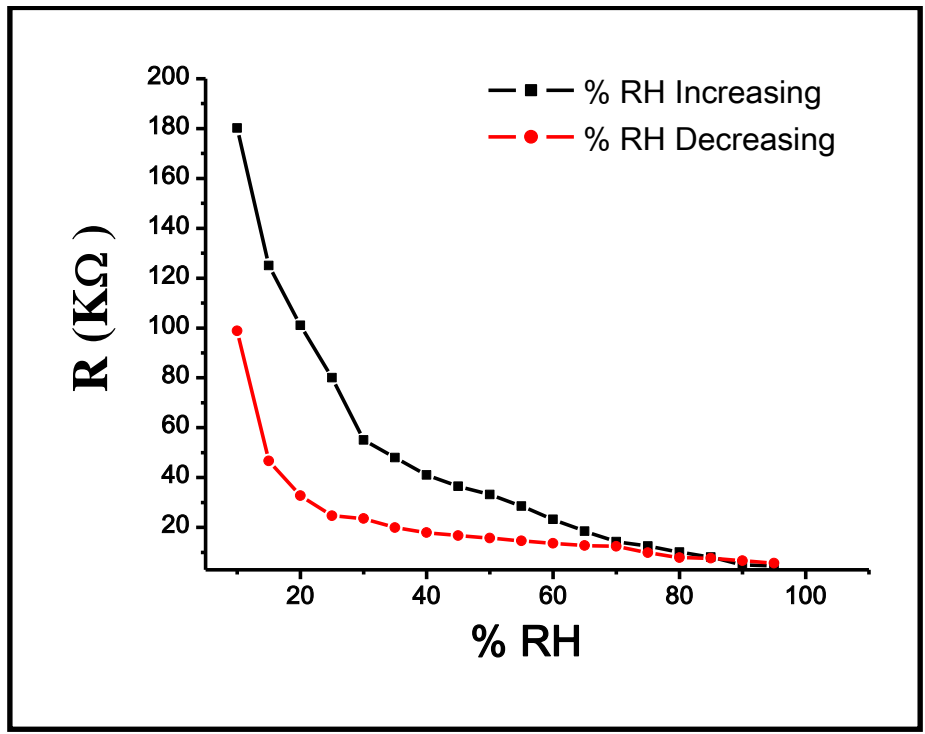

**Figure 8** Hysteresis behaviour of Sn$_{.918}$Sb$_{.109}$O$_2$ pellet annealed at 600ºC.

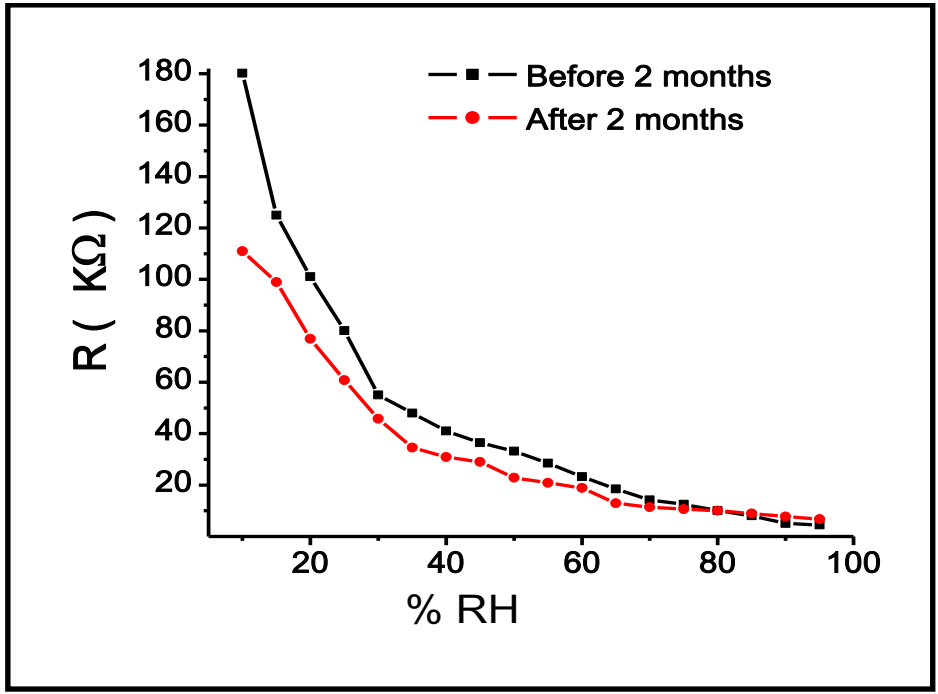

**Figure 9** Ageing effect of $Sn_{.918}Sb_{.109}O_2$ pellet annealed at 600ºC.

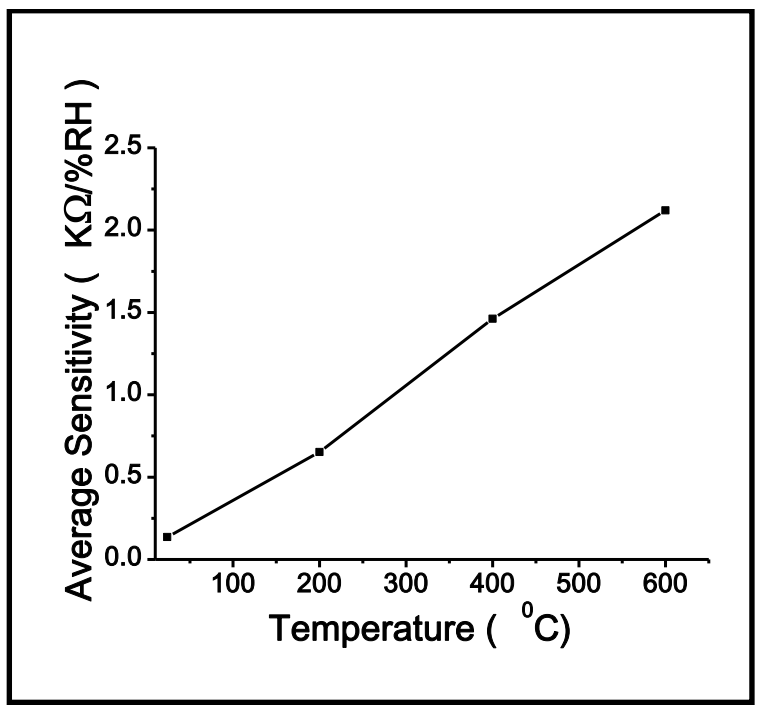

**Figure 10** Variations in average sensitivity with different temperatures.